\documentclass{iopart} %
\usepackage{iopams}  
\usepackage{graphicx,graphics}	
\graphicspath{{images/},{figures/}}
\usepackage[utf8]{inputenc}
\usepackage{eqnarray}
\expandafter\let\csname equation*\endcsname\relax 
\expandafter\let\csname endequation*\endcsname\relax
\usepackage{amsmath,amssymb,bbm}

\graphicspath{{images/},{figures/},{eps/}}

\usepackage{amsmath, amsthm, amssymb}

\newcommand{\pder}[2]{\frac{\partial#1}{\partial#2}}

\def\>{\rangle} \def\<{\langle}
\def\q{{\textrm{q} }}
\def\env{{\textrm{env} }}
\def\cen{{\textrm{cen } }}

\newcommand{\mcH}{\mathcal{H}}
\newcommand{\mcP}{\mathcal{P}}

\begin{document}
\title{Random density matrices versus random evolution of open system}
\newcommand{\unam}{Universidad Nacional Aut\'onoma de M\'exico, M\'exico}

\renewcommand{\paragraph}[1]{{\it #1.--}}

\newcommand{\ifunam}{Instituto de F\'{\i}sica, \unam}
\newcommand{\ifunamint}{\address{Instituto de F\'{\i}sica, Universidad Nacional Aut\'onoma de M\'exico, 01000 M\'exico D.F., Mexico}}

\newcommand{\cic}{\address{Centro Internacional de Ciencias A. C., Avenida Universidad s/n, 62131 Cuernavaca,
Mexico}}
\newcommand{\icf}{\address{Instituto de Ciencias F\'{\i}sicas, 
Universidad Nacional Aut\'onoma de M\'exico, Avenida Universidad s/n, 62210 Cuernavaca, Morelos, Mexico}}

\author{Carlos Pineda} \address{carlospgmat03@gmail.com} \ifunamint
\author{Thomas H. Seligman} \icf \cic
 
\begin{abstract}
We present and compare two families of ensembles of random density matrices.  The first, static ensemble, is obtained foliating an unbiased ensemble of density matrices. As criterion we use fixed purity as the simplest example of a useful convex function.  The second, dynamic ensemble, is inspired in random matrix models for decoherence where one evolves a separable pure state with a random Hamiltonian until a given value of purity in the central system is achieved.  Several families of Hamiltonians, adequate for different physical situations, are studied.  We focus on a two qubit central system, and obtain exact expressions for the static case.  The ensemble displays a peak around Werner-like states, modulated by nodes on the degeneracies of the density matrices.  For moderate and strong interactions good agreement between the static and the dynamic ensembles is found. Even in a model where one qubit does not interact with the environment excellent agreement is found, but only if there is maximal entanglement with the interacting one.  The discussion is started recalling similar considerations for scattering theory. At the end, we comment on the reach of the results for other convex functions of the density matrix, and exemplify the situation with the von Neumann entropy. 
\end{abstract}
\pacs{05.30.Ch,03.65.-w,03.65.Yz}
\maketitle
\section{Introduction} 

Evolution of open quantum systems is a subject of deep importance, since the
early days of quantum mechanics~\cite{vN55a} and has gained practical importance
since the boom of quantum information~\cite{breuer2007theory}.  One is often
not interested in the evolution of the environment, and under very reasonable
assumptions, one can greatly simplify the
discussion~\cite{Lindblad1976,Gorini1976}.  Random matrix theory (RMT) is used
to describe statistical aspects of ergodic quantum systems, and more recently,
to provide a framework for the description of
decoherence~\cite{Gorin2003,2008NJPh...10k5016G}.  Yet the random matrix theory
of the states themselves, i.e. of the density matrix is still in its infancy. A
first approach is given in \cite{PhysRevLett.104.110501,Nadal2011}, where an
``unbiased'' ensemble of density matrices is constructed imposing on random
covariance matrices the condition that their trace must be one.  However
correlations are important. We propose to make the next step in this direction,
by giving a recipe how to impose a fixed purity on such an ensemble, analogous
to the introduction of microcanonical ensembles in statistical mechanics.
Purity is chosen here, exclusively because of its analytic simplicity, but the
development presented is by no means restricted to this quantity.  Towards the
end of the manuscript we discuss entropy as an alternate quantity to be fixed.
We shall focus on quantities that are known to measure decoherence or
entanglement.   We are in part guided by the direct construction of ensembles
of scattering matrices~\cite{Mello1985254} and both differences and analogies will be
discussed.
The use of purity allows some analytic considerations (not possible with more
complex quantities) after which a very detailed discussion of a four level
central system will follow. There we can illustrate the behavior of density
matrices of fixed purity in terms of simple plots that show prominent features.
We shall also compare the results extensively with those obtained by using
dynamical RMT models for unitary evolutions of the total system (central system
plus environment)~\cite{2008NJPh...10k5016G}, i.e. RMT models for the Hamiltonian in a more traditional
setting~\cite{haakebook, pinedalong}, which have delivered interesting insights
in the field of decoherence.  The models here considered range from the
bluntest description of an open quantum system~\cite{Gorin2003,
PhysRevLett.107.080404}, to models which take into account the internal
structure of the Hilbert space~\cite{pinedaRMTshort}, and for which a strong
dependence on the initial condition is discussed.

In section \ref{sec:micro} we shall outline possible approaches to our
problem and compare them with the situation known for random scattering
matrices, where a similar development occurred nearly 30 years ago. In section
\ref{sec:static} we explicitly show how to construct ensembles of fixed purity
obtaining analytic results in many cases. We study in detail how the structure
of the Jacobi determinants enters and why we should take it into account. In
the same section we specialize to a four degree of freedom central system in
order to illustrate our result. In section \ref{sec:rmtdynamical} 
numerical calculations are presented for the more
standard approach of using RMT for Hamiltonians and couplings and compare the
results with the ones from the random density matrix ensembles. A wide variety
of RMT dynamics are used and we thus gain insight into the effectiveness of the
new method proposed. 
The generality of our arguments is insinuated by the use of von Neumann entropy
as a criterion to establish classes of density matrices in section
\ref{sec:entropy}. Finally we reach some conclusions and give an
outlook.

\section{Microcanonical ensembles for random density matrices} 
\label{sec:micro}

We shall devote this section to a discussion about the ensembles to be worked,
the motivation for their introduction and also some alternatives. We start by
recalling  the random matrix descriptions of scattering, to contextualize the
situation.  A dynamic random matrix theory for the $S$-matrix was developed
(see e.g. ~\cite{Agassi1975145, VWZ}) and has met great success.  It is based
on the introduction of random Hamiltonians and channel couplings into standard
formulae for the $S$-matrix.  Later a static approach, namely the direct
construction of a random ensemble of $S$-matrices was developed~\cite{SCali,
Mello1985254}, though the dynamical approach remains dominant in the
field.  As we propose to develop a ensemble of random density matrices as an
alternative to the existing dynamical description in terms or random
Hamiltonians this back flash will prove helpful and we shall emphasize
analogies and differences.

Both for the scattering matrices and for density matrices a simple ensemble for
what we may consider an appropriate a priori was/is previously known. For
scattering problems this would be the unitary or unitary symmetric matrices
depending on whether we consider time reversal breaking or conserving
situations. Both are so-called ``circular'' versions of Cartan's classical
ensembles \cite{cartanRMT, mehta}.  For density matrices
Nadal and Majumdar have recently proposed an ensemble~\cite{Nadal2011}. 
These
ensembles in both cases are very useful as a starting point, but they need
some refinement to describe a wider set of physical situations, because in the
scattering case they describe ``total absorption'', which is not very realistic
and in the density matrix case equipartition, which is not very interesting.

For scattering problems the simplest relevant case is the one where an
``optical'' $S$-matrix describes direct processes which vary slowly as a
function of energy (and are taken as energy independent) and a ``compound''
part in the spirit of Niels Bohr~\cite{bohr1939mechanism} which fluctuates fast implying
long times.  In this case the former part is assumed known, and usually easy to
measure and/or to evaluate, while the latter is represented by a random matrix
in the dynamical ansatz.  The construction of an appropriate ensemble produces
the expected averaged $S$-matrix describing this slowly varying part often
known as the optical $S$-matrix. On one hand the average $S$-matrix is not
really a unitary scattering matrix; it is actually sub-unitary.  On the other
hand an average density matrix will always again be a density matrix with all
its properties.  This matter is complicated by the fact, that the density
matrix itself defines an ensemble of quantum systems as well as providing a
description of a subsystem of a larger system This is the first important
difference, because in the case of the scattering matrix we explicitly seek an
ensemble of $S$-matrices fluctuating around a mean value which is not a unitary
scattering matrix. 
As an average over density matrices is in turn a density matrix, 
fixing this average leads to a trivial ensemble. 
In this paper we will concentrate on fixing a single scalar, mainly
purity, and construct the ensemble for its value.

As for the case of the $S$-matrix previous efforts exist to work with random
matrices in a dynamical way, but the above noted difference leads to
different approaches. Early work used an ensemble of Hamiltonians to evolve the
total system and then calculate the evolution of purity (and also concurrence
for a central system constituted by two qubits)  obtained from the evolution of
an initial product state \cite{pinedaRMTshort, pineda:012305} while more recent
work calculates the average density matrix directly
\cite{gorin2008random,MGS2015} and then obtains the average value say of
purity for that corresponding density matrix. Note that both are dynamical
approaches, but for quantities non-linear in the density matrix, there will
be a difference in the result. In practical examples this difference seems
small, and it seems reasonable to compare the static approach to the dynamical
calculation of density matrices as we will indeed do.

Constructing ensembles with certain restrictions is usually done in one of two
ways: The more notorious way is the use of Jaynes principle \cite{jaynes1,
jaynes2}, where we start form an {\it a priori} probability distribution and
find the distribution that minimizes the information, using Lagrange
multipliers, while  fulfilling other conditions.  These averages (or
expectation values in a quantum context) may result from experiments or
theoretical considerations~\cite{rlevine87:dynamics}.  The other alternative is
to impose the restriction strictly on every element of the ensemble
thus creating a ``microcanonical'' ensemble. 
In the case of scattering matrices
the latter approach is impossible, because the optical
$S$-matrix is not
a unitary scattering matrix, and we wish all members of the ensemble to be
such. Historically the first non-trivial ensembles of scattering matrices where
introduced using Jaynes principle and fixing the average $S$-matrix to lowest
order \cite{Mello1985254, SM2Nuclphys, SM31chann}. Later analytic properties of
the $S$-matrix demonstrated that the Poisson kernel multiplied by the measure
of Cartan's ensembles leads to a measure with any desired fixed average $S$-matrix.
More general information-theoretical arguments show that this is in
some sense the minimal information solution among a family of ensembles that
yield the same average $S$-matrix \cite{SCali, Mello1985254}. This
result is used extensively (see e.g.  \cite{baranger1994mesoscopic,
brouwer1995generalized, kuhl2005direct}) and yields, where applicable, similar
results to the dynamical model. 

For the density matrices with fixed average purity we have more freedom.  We
can choose the micro-canonical ensemble, where every member has the same purity
or a canonical ensemble, where purity has a given average value.  We have
chosen the former because the original definition of the unbiased ensemble
\cite{Nadal2011} directly includes unit trace as a delta function, and thus it
seemed technically easier to include fixed purity the same way.  We do not
expect a significant difference between the canonical and the micro-canonical
approach, but will not explore the conditions under which this is the case in the
present paper.


Existing static models have a limitation as time (or energy) have not been
included at this point: correlations between different times or different
energies (frequencies) cannot be obtained. On the other hand static ensembles
provide a very direct insight as to which properties govern the behavior of the
system beyond obvious things like phase space volume or noise.

\section{Eigenvalue distribution at fixed decoherence} 
\label{sec:static}
\subsection{General considerations} 

Consider the reduced density
matrix $\rho$ acting in our central system, resulting from taking the partial
trace of a random pure state $|\psi\>$, according to the Haar measure of the
unitary group from a larger space composed of our central system, and 
an auxiliary environment~\cite{wishartoriginal}. Namely, we take
%
\begin{equation}
\rho = \tr_{\mcH_\env} |\psi\>\<\psi|,\, \quad 
|\psi\>\in \mcH
\label{eq:statelarger}
\end{equation}
with 
\begin{equation}
\mcH=  \mcH_\env \otimes \mcH_\cen{},
\label{eq:hilbert:space:structure}
\end{equation}
and set $n=\dim \mcH_\cen{}$, and $m= \dim \mcH_\env$.
Notice that this {\it ensemble} of density matrices coincides with a Wishart
ensemble, normalize to unit trace. 
The density matrix $\rho$ has
eigenvalues $\lambda_i$, $i=1,\dots,n$ that (i) are connected by the
normalization condition 
\begin{equation}
\sum_i \lambda_i=1,
\label{eq:normalization}
\end{equation}
and (ii) are required to be non-negative:
\begin{equation}
\lambda_i \ge 0, \quad \forall i.
\label{eq:semipositive}
\end{equation}

Within this ensemble, we shall consider manifolds of codimension one, resulting
from fixing a quantity that depends only on the eigenvalues of the reduced
density matrix. This is inspired by thinking about a system with a {\it fixed
degree of decoherence}, as measured by purity. 
%
Purity is defined as 
\begin{equation}
P=\tr \rho^2
=  \sum_{i} \lambda_i^2.
\label{eq:defpurity}
\end{equation}
At this point it must be remembered that purity is just one of
infinitely many convex functions that can be used to characterize 
decoherence; its main advantage is its simple analytic
structure.  We shall later also consider the von Neumann entropy due to its
information theoretical meaning and general popularity. It is good to
remember that additivity, which is the main advantage of entropy among convex
functions, seems meaningless in the context of entanglement. Handling of the
full set of convex function and a use of the partial order implied, seems
unrealistically complicated despite of the availability of results for all
R\'enyi entropies~\cite{Nadal2011}.  Following~\cite{Nadal2011} one finds that
the distribution of the eigenvalues of the reduced state $\rho$ is given by
\begin{equation}
\mcP(\vec \lambda) \propto
\delta\left( \sum_i \lambda_i -1 \right)
\delta\left( \sum_i \lambda_i^2 -P \right)
\prod_i \lambda_i^{|m-n|}
\prod_{i<j} (\lambda_i-\lambda_j)^2
\label{eq:distr}
\end{equation}
where a term that accounts for the restriction to fixed purity is included. 

Up to unitary transformations in the central system, 
there are $n$ parameters characterizing our state
(the real eigenvalues of the $n$-dimensional
density matrix $\rho$), constrained with a physicality
condition (trace equal to 1) and the additional fixed purity constrain, both
conditions being scalar. Thus a $n-2$ dimensional space of free parameters is 
left. We want to do a mapping from the eigenvalues of a density matrix to a
meaningful and minimal space which can  be plotted and thus have a deeper
understanding of the ideas to be developed. A simple idea is to choose the
first $n-2$ eigenvalues, and let the other two be determined by restrictions
\eref{eq:normalization} and \eref{eq:defpurity}.  Given that $s_1 =
\sum_{i=1}^{n-2}\lambda_i$ and $s_2 = \sum_{i=1}^{n-2}\lambda_i^2$, 
the other two eigenvalues are
\begin{equation}
\lambda_{n-1},\lambda_n = \frac{1-s_1 \pm \sqrt{2(P-s_2)-(1-s_1)^2}}{2}.
\label{eq:other:two}
\end{equation}
A term that accounts for the Jacobian of the transformation is needed to correctly 
transform the probability densities. 
The transformation will take
$\lambda_1,\dots,\lambda_n$ to 
$\lambda_1,\dots,\lambda_{n-2}$, $\sum_{i=1}^{n}\lambda_i^2$, and
$\sum_{i=1}^{n}\lambda_i$.
That said, the determinant of the Jacobian has only a nontrivial contribution, given 
by a $2\times2$ block:
\begin{equation}
J=\begin{vmatrix} 1& 1 \\-2\lambda_{n-1} & - 2\lambda_{n}\end{vmatrix}=2 \sqrt{2(P-s_2)-(1-s_1)^2}.
\end{equation}

\begin{figure} 
\begin{center}
\includegraphics[width=\columnwidth]{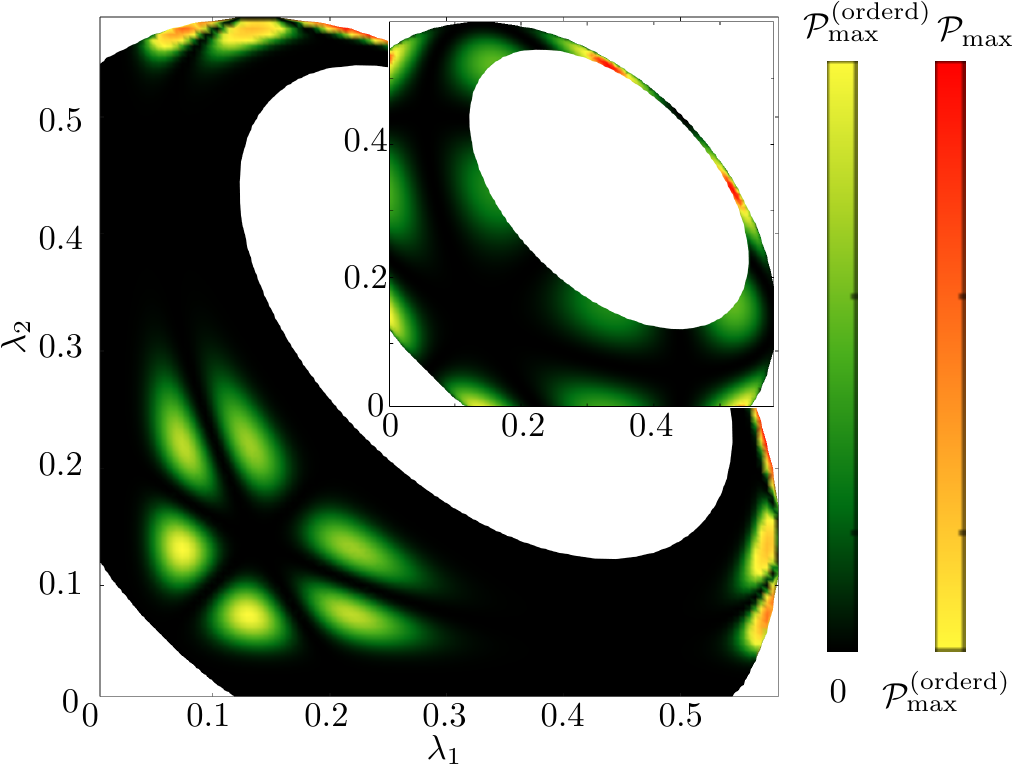}
\end{center}
\caption{\label{fig:example:smalldims} (Color online)  Probability distribution
of 2 eigenvalues for $n=4$, at fixed purity $P=0.4$ for $m=16$ (main figure) and $m=4$ (inset), in
arbitrary units. We have supplied two color scales. On green/yellow, the
maximum correspond to the maximum when the eigenvalues are ordered. On
yellow/red when we allow arbitrary order. This allows to observe better the details. 
The white region leads to nonphysical eigenvalues. For $m=4$, the probability
tends to the axes, whereas for larger $m$s, it has zero probability density
there. }
\end{figure} 

We can visualize the distribution for $n=4$ in
\fref{fig:example:smalldims}.
This distribution has a peak at $n-1$ degenerate states,
modulated by a repulsion of levels, which, altogether creates
a ``cleaved peak''.  Notice that at $m=n$ there is a qualitative
change of behavior, and a repulsion from the planes $\lambda_i=0$ is not observed,
since the term $\prod_i \lambda_i^{|m-n|}$ yields a constant.
Level repulsion, however, is always present.

In order to gain some insight into the behavior of the eigenvalue distribution, 
and to make an approximation for large environments, 
let us first rewrite~\eref{eq:distr} as
\begin{equation*}
\mcP = C_{m,P} J g^{|m-n|}(\lambda) V(\lambda),
\end{equation*}
with $\lambda$ being the vector that groups all $\lambda_i$ and 
$C_{m,P}$ a normalization constant.  The function 
\begin{equation}
g(\lambda) = g(\lambda_{1,\dots,n-2},P) = \prod_{i=1}^n \lambda_i 
\label{eq:g}
\end{equation}
is the product of all eigenvalues, but can be regarded as a function of the
first $n-2$ eigenvalues and the purity.
Finally, 
\begin{equation}
V(\lambda) = V(\lambda_{1,\dots,n-2},P) = 
\prod_{i<j} (\lambda_i-\lambda_j)^2
\label{eq:vandermonde}
\end{equation}
is a Vandermonde determinant. These two terms, $g$ and $V$, 
shall be analyzed separately. 
We will see that $g(\lambda)$ is responsible for a peak around a
$n-1$ degenerate state, whereas the determinant is responsible for the nodes
that appear modulating such peak. \par
\paragraph{The peak} 
We find explicitly that 
\begin{equation}
g(\lambda) =
\frac{
s_2 
+\left( s_1 \right)^2 
-2 s_1 
+1-P
}{2}
\prod_{i=1}^{n-2} \lambda_i .
\label{eq:defgtwo}
\end{equation}
To obtain the extrema of the function, the partial derivatives are
set to zero. Assuming $\lambda_i >0$ and rearranging, we obtain from 
the subtraction of the equations obtained from the partial derivative
of $g(\lambda)$ with 
respect to $\lambda_j$ and $\lambda_k$:
\begin{equation}
(\lambda_j - \lambda_k)
\left(\lambda_j + \lambda_k + s_1 -1\right)=0
\label{eq:condition:one}
\end{equation}
and from the sum of all partial derivatives, 
\begin{equation}
s_2 
+\left( s_1 \right)^2
+2\frac{1-n}{n} s_1 
+\frac{n-2}{n}(1-P)
=0.
\label{eq:condition:two}
\end{equation}
In general, these equations have an exponentially large number of solutions, not all of them 
physical, but a detailed analysis
turns out to be cumbersome. We first focus on the particular case
in which all $\lambda_i=\Lambda$, $i=1,\dots,n-2$, which clearly 
leads to a solution of all eqs.~(\ref{eq:condition:one}). From 
\eref{eq:condition:two}, and solving a simple quadratic equation, 
one obtains two solutions, 
\begin{equation}
\Lambda_{\pm} = \frac{1}{n} \pm \frac{1}{n}\sqrt{\frac{nP-1}{n-1}}. 
\label{eq:lambdapm}
\end{equation}
Each of this highly degenerate eigenvalues determines the other two eigenvalues, 
according to \eref{eq:other:two}. These
are $\lambda_{n-1}=\Lambda_{\pm}$ and 
$\lambda_{n}=1-(n-1)\Lambda_{\pm}$. This corresponds to a state  that is a mixture
of a pure state and the maximally mixed state. Notice that for 
$P >1/(n-1)$, $\Lambda_+$ results in a negative (and thus inadmissible)
$\lambda_{n}$.

We can simplify the expression for the probability density
near  the maximum corresponding to the point
$\lambda_i = \Lambda_-$, $i=1,\dots,n-1$. We expand in Taylor 
series around the maximum, and note that
\begin{align}
\pder{^2 g}{\lambda_j^2}{\biggr |}_{\lambda_i=\Lambda_-}
= 2 \pder{^2 g}{\lambda_j \lambda_{k\ne j}}{\biggr |}_{\lambda_i=\Lambda_-}
  = 2 \Lambda_-^{n-2}\left( n-\frac{1}{\Lambda_-} \right),
\label{eq:partial:derivatives}
\end{align}
with $j=1,\dots,n-2$. This leads to 
\begin{equation}
g(\lambda_{1,\dots,n-2}, P) \approx
g_\text{max} - \alpha 
\tilde \lambda
\cdot A \cdot
\tilde \lambda^T
\label{eq:approxpoly}
\end{equation}
with 
$$A=
\begin{pmatrix} 
2 & 1 & \dots & 1 \\ 
1 & 2 & \dots & 1 \\
\vdots & \vdots & \ddots & \vdots \\
1 & 1 & \dots & 2 
\end{pmatrix},$$
$\alpha=
\Lambda_-^{n-2}\left( n-1/\Lambda_- \right)$, 
$\tilde\lambda = \begin{pmatrix} \lambda_1-\Lambda_- & \cdots &
  \lambda_{n-2}-\Lambda_- \end{pmatrix}$
and 
$g_\text{max}=g|_{\lambda_i=\Lambda_-}$.
Recall that 
$\rme^{-x} \approx 1-x$ for small values of $x$, so that, near a maximum,
one can approximate a polynomial with a Gaussian:
$ (1-x^2)^m \approx \rme ^{-mx^2}$ 
for $m\gg 1$ and $-1 < x < 1$. 
It is then found that 
\begin{equation}
g(\lambda_{1,\dots,n-2}, P)^{|m-n|} \approx
g_\text{max}^{|m-n|}
\exp \left(
-\frac{\tilde\lambda^T \cdot A \cdot \tilde\lambda}{\sigma_{m,P}^2}\right)
\label{eq:approximation:g}
\end{equation}
 with the average standard deviation given by
with the factor
$\sigma_{m,P}^2=g_\text{max}/\alpha(m-n)$
determining the standard deviation of each of the eigenvalues.\par
\paragraph{The  Vandermonde determinant} 
We will study the situations in which $V(\lambda)=0$, as this will give information 
about the nodes of the probability density that will help understand its behavior. 
This happens when at least one of the terms $\lambda_i - \lambda_j=0$ ($i\ne j$). 
In the space of the first $n-2$ eigenvalues, the conditions
\begin{align}
\lambda_1=\lambda_2,\quad  & \lambda_1=\lambda_3, & \dots, \quad &\lambda_1=\lambda_{n-2} \nonumber\\
                     & \lambda_2=\lambda_3,       &\dots,  \quad &\lambda_2=\lambda_{n-2} \nonumber\\
                     &                      & \ddots             &                        \nonumber\\
                     &                      &         & \lambda_{n-3}=\lambda_{n-2} 
\end{align}
are simple hyperplanes. On the other hand, replacing 
the conditions $\lambda_j = \lambda_{n-1}$
and $\lambda_j=\lambda_n$ on 
\eref{eq:other:two} and squaring the discriminant
for $j=1,\dots,n-2$, define the $n-2$ different quadratic 
forms 
\begin{equation}
Q_j(\lambda_i) = \left( 2\lambda_j + s_1 -1\right)^2 - 2(P-s_2)+ (1-s_1)^2=0.
\label{eq:all:ellipsoids}
\end{equation}
All these curves are ellipsoids, as can be noted by the fact that all 
$\lambda$'s are bounded in these degree 2 polynomials. 
The condition $\lambda_{n-1}-\lambda_n=0$
can be calculated directly form \eref{eq:other:two}, and
leads to the ellipsoid 
\begin{equation}
Q_n = 2(P-s_2)-(1-s_1)^2 = 0,
\label{eq:physicalellipsoid}
\end{equation}
which marks the limits for which the expression~\eref{eq:other:two} lead to
real, rather than complex and thus inadmissible, values for $\lambda_{n-1,n}$.
Finally, the condition $\lambda_{n-1}=0$ constitutes another constraint, which
can also be calculated from~\eref{eq:other:two}, and one obtains 
\begin{equation}
Q_{n-1} = (P-s_2)-(1-s_1)^2 = 0.
\label{eq:physicalellipsoid:other}
\end{equation}
Notice that the condition $\lambda_{n}=0$ is automatically taken care of, as
from \eref{eq:other:two}, $\lambda_n \ge \lambda_{n-1}$.
The ellipsoids are really symmetry planes, that acquire this skew form since
the last two eigenvalues are given a special status. One could focus the study
to one region defined by any side of each of the ellipsoids \eref{eq:all:ellipsoids}.

At the peak, there is an $n-1$ fold degeneracy of the eigenvalues. 
The right hand side of~\eref{eq:vandermonde} may be approximated, considering 
only linear terms with respect to $\lambda_j -\Lambda_-$. We obtain
from the linear approximation $q_j(\lambda_i)$ of $Q_j(\lambda_i)$
\begin{align}
q_j(\lambda_i) &= 
 (\lambda_j-\Lambda_-) \pder{Q_j}{\lambda_j}{\biggr |}_{\lambda_i=\Lambda_-} \nonumber 
 +\sum_{k\ne j} (\lambda_k-\Lambda_-) \pder{Q_j}{\lambda_k}{\biggr |}_{\lambda_i=\Lambda_-} \\
&=4(n\Lambda_--1)[\lambda_j+s_1-(n-1)\Lambda_-],
\label{eq:nodes:as:planes}
\end{align}
a series of $n-2$ hyperplanes. This approximation will be useful when combined with \eref{eq:approximation:g}
to yield a simple expression near the degeneracy.

\begin{figure} 
\begin{center}
\includegraphics[scale=1.3]{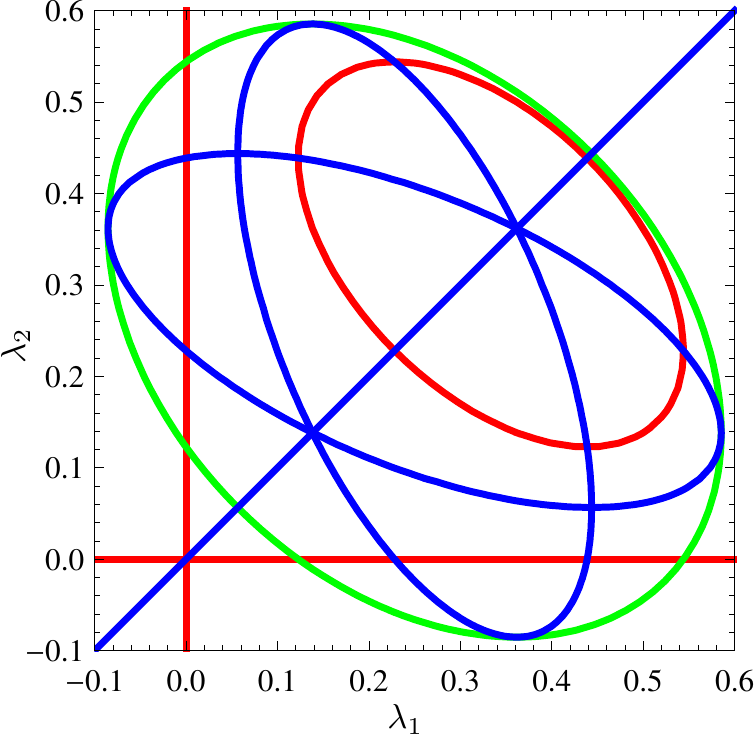}
\end{center}
\caption{\label{fig:manchatwoD} (Color online) Boundaries for the possible
physical states, and visualization of the Vandermonde determinant in the
$\lambda_1$, $\lambda_2$ plane, for $P=0.4$ and $n=4$.  The regions where the physical
density matrices live is above the horizontal [red] line ($\lambda_2\ge 0$), to
the right of the vertical [red] line ($\lambda_1\ge 0$), inside the outer
[green] ellipse [real values for $\lambda_3$ and $\lambda_4$, 
eqs.~\eref{eq:physicalellipsoid} and \eref{eq:qthreefour}] and outside the
smaller [red] ellipse [$\lambda_3\ge0$, eqs.~\eref{eq:physicalellipsoid:other}
and \eref{eq:qfour}]. The geometric place of the
$\lambda_1=\lambda_2$ degeneracy is the diagonal line [blue], of the
$\lambda_1=\lambda_3$ and $\lambda_1=\lambda_4$ degeneracy is the most vertical
[blue] ellipse and the $\lambda_2=\lambda_3$ and $\lambda_2=\lambda_4$
degeneracy is the most horizontal [blue] ellipse, see
eqs.~\eref{eq:all:ellipsoids} and \eref{eq:qotf}.  } \label{fig:symmetries}
\end{figure} 
\subsection{The special case of $n=4$} 
\label{sec:four:fixed}
The case $n=4$ deserves special attention, as the numerics are carried on there, and 
a complete plot of the full distribution is possible.
The explicit form of $g$ and $\Lambda_-$ can be read directly from
eqs.~(\ref{eq:defgtwo}) and (\ref{eq:lambdapm}). 
The analysis of the conditions of degeneracy results more interesting. 
In this case, the ellipsoid 
corresponding to $\lambda_1=\lambda_3$ and $\lambda_1=\lambda_4$ is
\begin{equation}
Q_1(\lambda_1,\lambda_2)= 3\lambda_1^2+\lambda_2^2+2\lambda_1\lambda_2-2\lambda_1-\lambda_2+\frac{1-P}{2}.
\label{eq:qotf}
\end{equation}
This is an ellipse centered in $(1/4, 1/4)$ rotated an angle $\theta$ such that 
$\cot \theta=1+\sqrt2$ and with semi-axes of length $\sqrt{4P-1}/4\sqrt{1\pm2^{-1/2}}$.
$Q_2$ will be the  same up to an interchange of
$\lambda_{1}$ and $\lambda_2$. 
The border curve corresponding to $\lambda_3=\lambda_4$ 
leads to the polynomial  
\begin{equation}
Q_4(\lambda_1, \lambda_2)=3\lambda_1^2+ 3\lambda_2^2+ 2\lambda_1\lambda_2-2\lambda_1-2\lambda_2-2P+1.
\label{eq:qthreefour}
\end{equation}
This defines an ellipse with center in  $(1/4, 1/4)$, rotated by $\pi/4$ 
and with semiaxis  $(\sqrt{4P-1}/3, \sqrt{(4P-1)/3})$.

Semi-definite positivity of the density operator is guarantied
if one restricts to the area delimited by $\lambda_i \ge 0$, which amounts to
consider the quadrant $\lambda_{1, 2}\ge 0$, and $\lambda_3 \ge 0$
as, by construction, $\lambda_4\ge\lambda_3$. The condition 
$\lambda_3 \ge 0$ is met in the exterior of the ellipsoid defined by 
$Q_3(\lambda_1, \lambda_2)=0$ with
\begin{equation}
Q_3(x, y)=2x^2+ 2y^2 +2xy-2x-2y+1-P.
\label{eq:qfour}
\end{equation}
This is an ellipsoid centered in $(1/3, 1/3)$, rotated $\pi/4$ and with semiaxis 
$(\sqrt{3P-1}/3, \sqrt{(3P-1)/3})$. For $P<1/3$ this ellipse does not exist; as long 
as the previous conditions are met, $\lambda_4$ will be real. 

Using approximation \eref{eq:nodes:as:planes} the following simplified distribution is
obtained
\begin{equation}
\mcP(\lambda_1, \lambda_2) = C_P
\exp\left[
-\frac{\tilde\lambda^T \cdot A \cdot \tilde\lambda}{\sigma_{m,P}^2}\right]
\left[ 
  (\lambda_1-\lambda_2)(2\lambda_1+\lambda_2-3\Lambda_-)(\lambda_1+2\lambda_2-3\Lambda_-) \right]^2,
\label{eq:approx}
\end{equation}
valid for large dimensions of the environment and sufficiently close to 
$\lambda_{1,2,3} \approx \Lambda_-$.
It is now simple to read the behavior of the function. It is indeed a 
6-fold peaked function arising from a Gaussian of width diminishing 
as $1/\sqrt{m}$, modulated by some parabolas touching zero. In this
approximation the center of the Gaussian is
located in the point $\vec \Lambda_-=(\Lambda_-, \Lambda_-)$ with
$\Lambda_-=(3-\sqrt{12P-3})/12$; a point corresponding to a triply 
degenerate state. The other peaks are in $\vec \Lambda_- + 
\sigma_{m,P} \vec v_i$ where $ \vec v_{1,\ldots,6} \in \{(\pm \sqrt3, 0),
(0,\pm \sqrt3, 0), \sqrt2(\pm1, \mp1) \}$. The last thing that should be
calculated are single variable distributions, 
which are presented in appendix~\ref{sec:distribution:smallest:eigenvalues}.

A glimpse at the complexity of the surface can be obtain analyzing the
stationary points of the surface. In this case, \eref{eq:condition:one} and
\eref{eq:condition:two} lead to four solutions. The first one corresponds to
$\lambda_1 = \lambda_2 = \Lambda_-$.  The other two eigenvalues are $
\Lambda_-$ and $1-3\Lambda_-$. A second solution, corresponding to triply
degenerate $\Lambda_+$ remains physical for $P\le 1/2$, after which the points
migrate to the unphysical region corresponding to the interior of the red
ellipse in \fref{fig:manchatwoD}. The other two solutions correspond to saddle
points of the function $g$, but for $P > 1/2$ also correspond to unphysical
eigenvalues.
There are additional maximums that are not detected by conditions
\eref{eq:condition:one} and \eref{eq:condition:two}, which live in the edge of
the domain, and thus would have to be calculated separately. These also
correspond to the same triply degenerate states with $ \Lambda_-$, but
reordered and are located at the points in which the blue and green ellipses 
in \fref{fig:manchatwoD} touch each other.

All these boundaries are plotted in \fref{fig:manchatwoD}, where one can visualize
the roll of  each of the conditions earlier discussed. The global effect can 
also be nicely seen in \fref{fig:example:smalldims}.

\section{Random states and random dynamics} \label{sec:rmtdynamical} 
We now wish to compare the results obtained for the static model discussed in 
the previous section with 
those of previously studied dynamical random matrix
models, in which decoherence of a central system can be
studied~\cite{1464-4266-4-4-325, Gorin2003, PhysRevLett.107.080404}.  These
models are time independent,
but will be used to evolve an initial state until a 
time in which the purity (or any other quantity of interest) reaches
the particular value of interest, thus obtaining an equivalent ensemble
of density matrices.

That said, consider again a Hilbert space with the structure 
${\mathcal H}\ =\ \mcH_\cen \otimes \mcH_\env $ [recall
\eref{eq:hilbert:space:structure}]. As in sec.~\ref{sec:four:fixed}, we also
restrict to the case in which $\mcH_\cen$ is a  4-level system. 
Special consideration will be given to the case 
in which $\mcH_\cen$ is composed of two
qubits; that is, when 
\begin{equation}
\mcH_\cen = {\mathcal H}_{\q_1} \otimes {\mathcal H}_{\q_2}.
\label{eq:two:qubits}
\end{equation}
where ${\mathcal H}_{\q_i} $ denote
qubit spaces.
In this space, unitary dynamics will be generated by random Hamiltonians, 
to be detailed later. Non-unitary dynamics
in the two qubit central system is thus induced by the unitary dynamics 
in the whole space, plus partial tracing over $\mcH_\cen$. 
%
As initial states, pure separable states in the whole Hilbert 
space are used:
\begin{equation}
  \label{eq:initialconditiongeneral}
  |\psi(0)\>=|\psi_\cen\>|\psi_\env\>,\quad       
|\psi_\cen\>\in
   \mcH_\cen,\,|\psi_\env\>\in \mcH_\env; .
\end{equation}
We take $|\psi_\env\>$ to be a random state, according to the Haar measure of
the unitary group in the environment, but $|\psi_\cen\>$ is to be chosen
according to the particular case under
study. \par
Kolmogorov distance will be used to quantify the difference between two 
distributions. Given two functions, $f(x)$ and $g(x)$, defined on 
a domain $X$, it is
\begin{equation}
K(f, g) = \frac12 \int_X |f(x)-g(x)| \rmd x.
\label{eq:kolmogorov}
\end{equation}
Notice that since we are dealing with distributions, the normalization condition
for a distribution $\mcP(x)$ is stated as $K(\mcP, 0)=\frac12$, and thus the distance
between two non overlapping (i.e. totally distinct) distributions $\mcP_1$ and
$\mcP_2$ is always 1. \par
\subsection{Global Hamiltonian} 
\begin{figure}
\begin{center}
\includegraphics[scale=1.3]{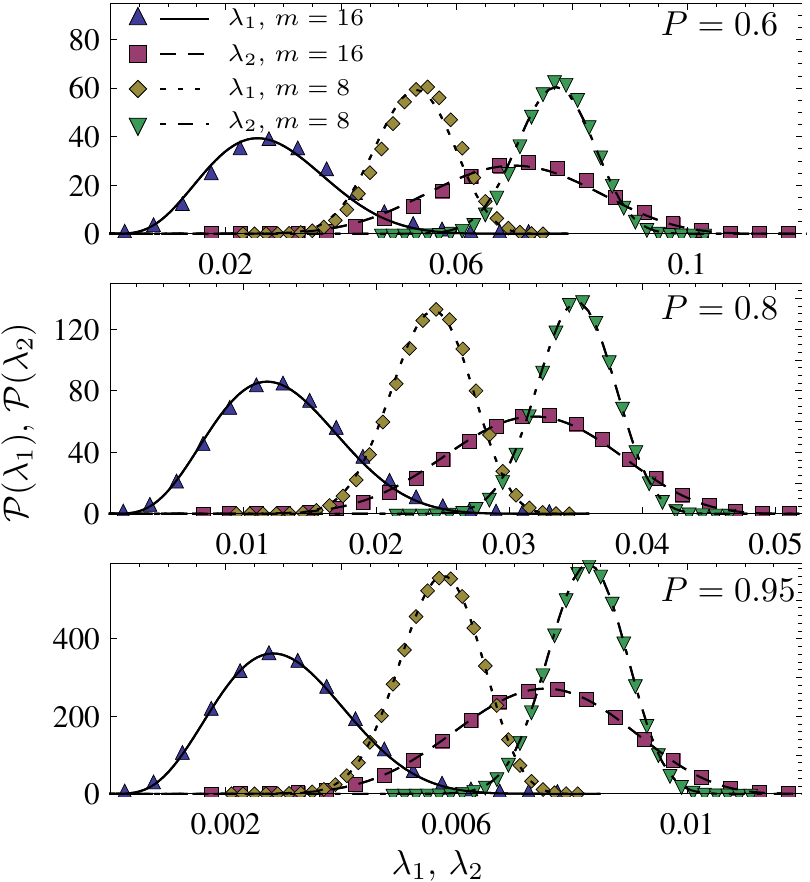}
\end{center}
\caption{\label{fig:global:hamiltonian} 
(Color online) 
Marginal distribution of the smallest two eigenvalues ($\lambda_1 \le \lambda_2$). 
The points are obtained with nonunitary evolution 
under the global Hamiltonian~\eref{eq:global:hamiltonian} with close to $10^6$
realizations, and the curves are obtained with the marginal distributions
\eref{eq:lambdaone} and \eref{eq:lambdatwo}, obtained by proper integration of
\eref{eq:distr}. We show different target purities and different environment
dimensions. }
\end{figure}

The simplest candidate for a random matrix model that describes decoherence is
simply choosing a Hamiltonian from one of the classical
ensembles~\cite{cartanRMT} that acts on the whole Hilbert space, i.e. on both
the central system and the environment~\cite{Gorin2003}. We shall call this
family {\it global Hamiltonian}.  These Hamiltonians are attractive because of
their analytic simplicity~\cite{PhysRevLett.107.080404}.  They also model the
strongest interaction between central system and environment. We shall choose
to pick the Hamiltonian from the GUE (Gaussian unitary ensemble) as it is the
simplest ensemble, from an analytical point of view. 

For this case, we can write simply 
\begin{equation}
H = H^{(\text{GUE})}_{\env,\cen}
\label{eq:global:hamiltonian}
\end{equation}
where the superindex indicates the ensemble from which the operator is chosen, and the subindices
indicate the spaces in which they act non-trivially. For this case, and due to
the invariance of the GUE, one can choose the initial state of the central
system arbitrarily, with no effect on the results regarding
the eigenvalue density of the evolved state of the central system. 

The eigenvalue density of the density matrices produced by the non-unitary
dynamics induced by the Hamiltonian \eref{eq:global:hamiltonian} are similar
to those for the static situation, \eref{eq:distr}. However, 
differences can be observed by inspection, see \fref{fig:global:hamiltonian}.
These differences are larger for smaller purities, and seem to be independent
of the size of the environment. However, even for an intermediate purity of
$P=0.8$, it is difficult to see any differences.
If one is not interested in very precise data, or only in high purity,
quantitative information can be extracted from the static model regarding
the evolved state.

\begin{figure} 
\begin{center}
\includegraphics[scale=1.3]{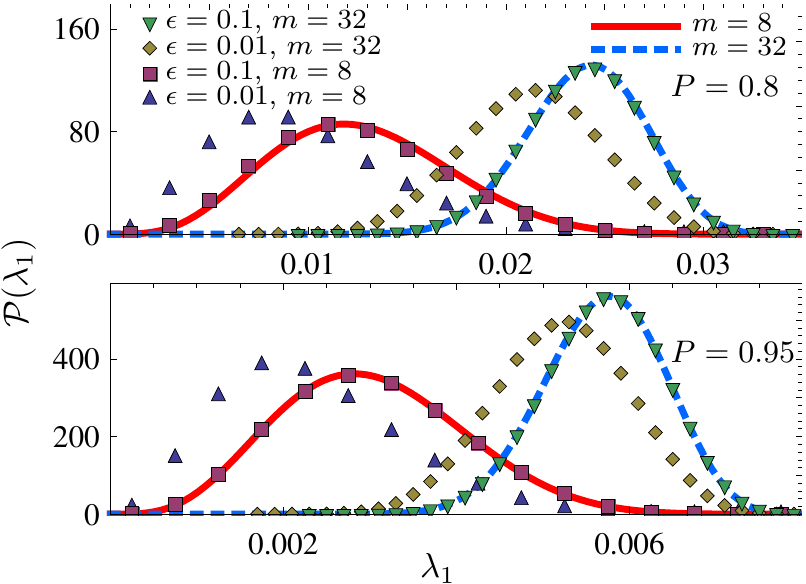}
\end{center}
\caption{\label{fig:coupling:hamiltonian} 
Marginal distribution of the smallest eigenvalue $\lambda_1$. 
The points are obtained with nonunitary evolution 
under the tunable coupling Hamiltonian~\eref{eq:weak:four:hamiltonian} with close to $10^5$
realizations, and the curves are obtained with the marginal distributions
\eref{eq:lambdaone} and \eref{eq:lambdatwo}, obtained by proper integration of
\eref{eq:distr}. We show two different target purities, different environment
dimensions and different couplings. For very small couplings, the ensemble
is not well approximated by \eref{eq:distr}, but already for moderate couplings, 
the ansatz is very good.}
\end{figure} 
\subsection{Tunable coupling}
Hamiltonian \eref{eq:global:hamiltonian} couples the central system and 
its environment. However it does not take into account the structure of 
Hilbert space~\eref{eq:hilbert:space:structure}. One way to do that is 
to provide the Hamiltonian governing the system with a similar 
tensor product structure. Moreover this will lead the idea of tunable
coupling which is very convenient when studying open quantum systems. 
We thus use 
\begin{equation}
H = H^{(\text{GUE})}_{\env} + \epsilon V^{(\text{GUE})}_{\env,\cen},
\label{eq:weak:four:hamiltonian}
\end{equation}
which will be called {\it tunable coupling Hamiltonian}, or
coupling Hamiltonian, for short. 
It must be noticed that the case
$\epsilon\to\infty$ corresponds, up to a rescaling in time, to the previous case, 
i.e. \eref{eq:global:hamiltonian}.

Regarding the eigenvalue density of the evolved state of the central 
system with 
Hamiltonian \eref{eq:weak:four:hamiltonian}, a significant 
dependency on the coupling strength was observed.   
Four regimes should be considered: weak coupling [for which 
$\epsilon = 0.01$ is taken as a representantive], intermediate coupling
[$\epsilon = 0.1$], strong coupling [$\epsilon = 1$], and 
very strong coupling [$\epsilon \to \infty$]. 
\par
For weak coupling there is a clear discrepancy, for all purities and dimensions
examined. The agreement does not seem to improve for larger dimensions,
see \fref{fig:coupling:hamiltonian}. It must be noticed that even though there
is a significant disagreement between the two ensembles, the nodes can still be
observed, and for a qualitative description, the static ensemble is still
useful. 
\par
For intermediate coupling $\epsilon=0.1$, the agreement is very good, and only
for smaller dimensions and purities can the difference be observed by
inspection, see \fref{fig:coupling:hamiltonian}. Here, it seems that both,
increasing the purity at fixed environment dimension, and increasing the
dimension at fixed purity leads to the same ensemble as the static one. For
strong coupling and very strong coupling, the results are indistinguishable
from the global Hamiltonian case, see \fref{fig:kolmogorov}.
\par
\subsection{Spectator Hamiltonian} 
An interesting variation of \eref{eq:weak:four:hamiltonian} has 
been studied in~\cite{pinedaRMTshort}. There, one studies a central system
whose Hilbert space is composed of two qubits. That is
when the Hilbert space is of the form \eref{eq:two:qubits}. 
There, the 
{\it spectator Hamiltonian} 
\begin{equation}
\label{eq:hamiltonianspectator}
H= H^{(\text{GUE})}_\env  +\epsilon V^{(\text{GUE})}_{\env,\q_1}
\end{equation}
was proposed where, again, the subindices indicate the part of the Hilbert 
space where they act non trivially.  
The first  term correspond to the free dynamics of the environment. The next 
represents the coupling of the first qubit to the 
environment.  Thus, the second qubit
is simply a \textit{spectator}, as  it has no coupling
to an environment.  
This is the
\textit{simplest} Hamiltonian for which we can analyze the effect of an
environment on a Bell pair \cite{pinedaRMTshort}.  
The environment Hamiltonian $H_\env$ will be
chosen from a classical ensemble~\cite{cartanRMT} of $m \times m$ matrices and
the coupling $V_{\env,\q_1}$ from one of $2m \times 2m$ matrices.   

\begin{figure} 
\begin{center}
\includegraphics[width=\columnwidth]{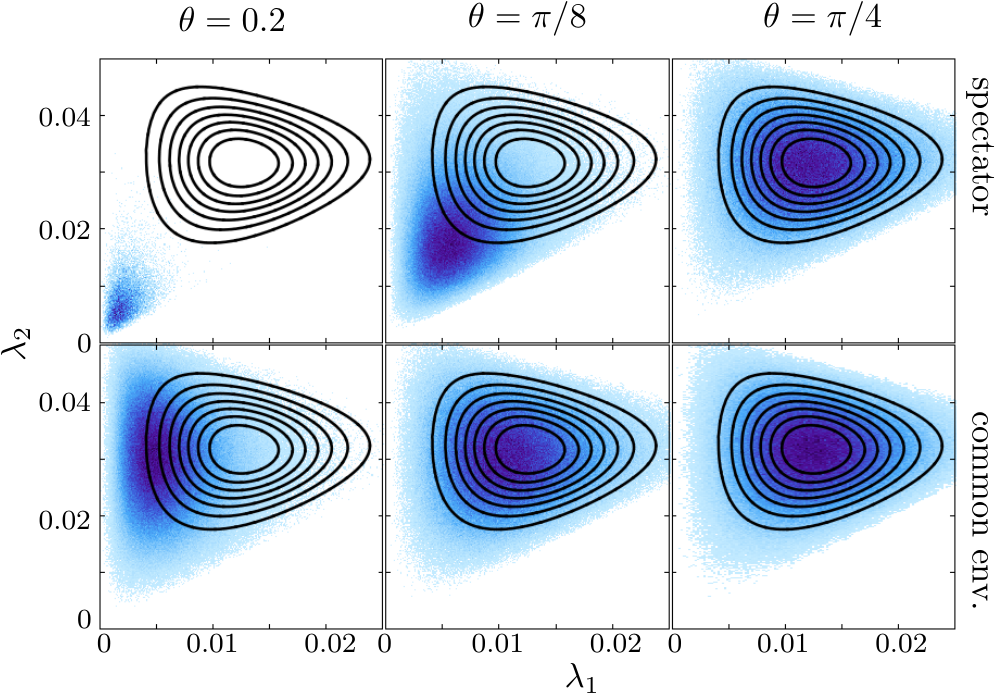}
\end{center}
\caption{Distribution of the two smallest eigenvalues for the spectator and the common
environment Hamiltonians eqs.~(\ref{eq:hamiltonianspectator}),
(\ref{eq:hamiltonian:common}),  
for various degrees of entanglement of the initial condition in 
the central system, modulated by 
$\theta$ in \eref{eq:general:two:qubit}. The target purity 
is, in this case, $P=0.8$ and we fix $m=8$. For the spectator case, low
entanglement tends to push the distribution towards the origin, and for 
the common environment, small entanglement tends to push the distribution 
towards the $y$ axis.  Both models are in good agreement with the
ansatz distribution for a maximally entangled initial state. The figure
shows a disymmetrized distribution, assuming non-decreasing eigenvalues, and
the scale is arbitrary. 
}
\label{fig:varyent}
\end{figure} 
This Hamiltonian is quite interesting, since it does not involve one of the
qubits. At least not from the dynamical point of view. This Hamiltonian adds a
structure to the problem: it creates the notion of a particle. One obvious new
ingredient when the Hilbert space is split is entanglement. In this case we
shall thus have a new parameter which is the entanglement of the initial state
in the two qubits. 

\begin{figure*}[ht] 
\begin{center}
\includegraphics[width=\textwidth]{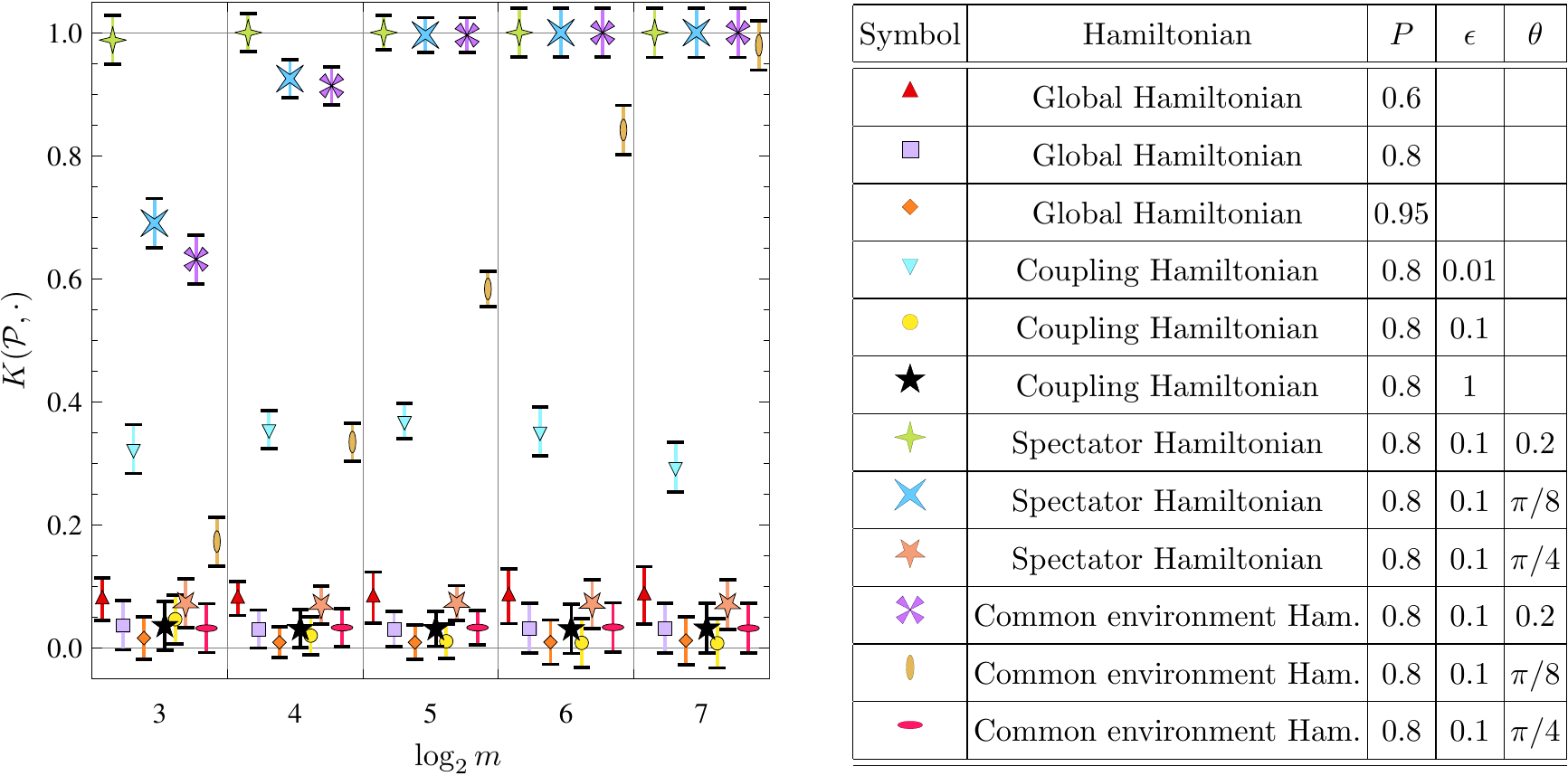}
\end{center}
\caption{We present a summary of the comparison between the static ensemble,
and the ensemble generated by random dynamics (left). 
In the ordinate $\log_2 m$ with only take integer values. The symbols representing the different
cases will appear for arbitrary values within the interval around the fixed integer value
of $\log_2 m$ to make then discernible. On the abscissa the Kolmogorov distance between
the dynamical model and the corresponding static one $K(\mathcal P, \cdot)$ is plotted. 
Good agreement between
both is observed for the global Hamiltonian, 
the coupling Hamiltonian for large enough coupling, and for the spectator and common environment Hamiltonian 
when the initial state has maximal entanglement. The symbols must be understood using the 
table on the right hand side. The error bars are obtained with the Kolmogorov  distance with 
respect to a Monte Carlo simulation with the same number of points as the one with 
the random dynamics, that is, $10^6$ points.  }
\label{fig:kolmogorov}
\end{figure*} 
For a totally disentangled initial state, the dynamics will produce in the 
two qubits a reduced state of the form
\begin{equation}
\rho(t) = \rho_1(t) \otimes |\psi\> \< \psi| 
\label{eq:separable:two:qubit}
\end{equation}
at all times.  Notice that the smaller eigenvalues $0$ are double degenerate.
The other two eigenvalues are totally determined by the normalization condition
and the fixed purity condition.

Taking into account the invariance of the ensemble defined by
\eref{eq:hamiltonianspectator} under local unitary operations, 
we notice that the initial states can be written with 
absolute generality as 
%
\begin{equation}
|\psi(0) \> = \sin \theta |00\> + \cos\theta |11\>,
    \quad \rho(0) = |\psi(0) \> \<\psi(0) |,
\label{eq:general:two:qubit}
\end{equation}
where $\theta$ is the only relevant parameter of the state, and
also modulates the initial entanglement. $\theta=0$ corresponds 
to an initially separable state, whereas $\theta=\pi/4$ to a
maximally entangled one.
If $\theta=0.2$ the situation is indeed close to the totally separable case.
In this case the distribution is totally different than 
expected from the static analysis, that is from \eref{eq:distr}, see
\fref{fig:varyent}.  However for a maximally entangled state $\theta = \pi/4$,
the results are similar as to the tunable coupling case with four levels.  That
is, acceptable agreement for very small coupling, and much better agreement
for intermediate coupling. Again, the agreement improves
with increasing purity, but does not seems to converge for increasing dimension of
the environment at a fixed purity, see \fref{fig:kolmogorov}. 
\par

\subsection{Common environment Hamiltonian} 
Even though Hamiltonian (\ref{eq:hamiltonianspectator}) is the simplest one
that allows us to study entanglement evolution~\cite{pinedaRMTshort}, in many
situations one would have coupling of all constituents of the
central system to the environment. To be closer to many experimental situations, 
consider what we call the {\it common environment} Hamiltonian
\begin{equation}
\label{eq:hamiltonian:common}
H= H^{(\text{GUE})}_\env  
+\epsilon V^{(\text{GUE})}_{\env,\q_1}
+\epsilon V^{(\text{GUE})}_{\env,\q_2}.
\end{equation}
Again, the indices indicate the subspaces in which they operate non-trivially,
with respect to \eref{eq:hilbert:space:structure} and \eref{eq:two:qubits}. 
This Hamiltonian, includes coupling of both qubits to an environment, but also 
takes into account a Hilbert space with the structure~\eref{eq:two:qubits}.

For the common environment Hamiltonian we have again a particular form of
the eigenvalues for an initially separable state. In this case, one can
approximate the dynamics as two channels acting independently on the two
qubits. Thus, neglecting correlations in the environment, for later times one
should have that
\begin{equation}
\rho(t) \approx \rho_1(t) \otimes \rho_2(t).
\label{eq:commong:two:qubit}
\end{equation}
%
One then has that the eigenvalues of the reduced density matrix are $(\lambda,
1-\lambda)\otimes (\lambda', 1-\lambda')$, which
gives an additional global restriction; the manifold in which the eigenvalues
lives has dimension one, instead of two as in the previous cases.
This will of course heavily influence the distributions of
eigenvalues. For a small amount of initial entanglement the situation should vary
continuously, and one indeed observes very significant differences with the
proposed ansatz, see \fref{fig:kolmogorov}.  For an initially maximally entangled state,
the results are again similar and are well reproduced by the static ansatz, see
figs.~(\ref{fig:varyent}) and (\ref{fig:kolmogorov}).

\subsection{A comparison using Kolmogorov distance} 
The results of this section can be summarized in a figure containing the
Kolmogorov distance between the results for the dynamical system, and the
statical ensemble, \eref{eq:distr}. We
present those results in \fref{fig:kolmogorov}.

Error bars, obtained using the Kolmogorov distance between the exact
distribution, and an equivalent Monte Carlo simulation [see
Appendix~\ref{sec:montecarlo}] with the same number of data points as the
dynamical situation, is also included.  This number should be interpreted
as the error arising both from finite sampling and finite binning.  

We can
see that the results for the static model and the dynamical models agree well
for the global Hamiltonian, and for the coupling Hamiltonian, as long as the
coupling is not too small.  If the coupling  is very small, there are
quantitative deviations, but the shape of the distribution remains similar. In
this case, one can see that there is a clear difference between the static and
the dynamical ensemble, which leads to intermediate values of $K$, very
different both from 0 and 1.  As coupling becomes very small, these deviations
increase.  Often such cases are associated with situations where the dynamical
model will actually not reach equipartition~\cite{PhysRevA.90.022107,
pinedalong}.  

For models in which structured coupling is present (spectator and
common environment models), one has similar results as for the tunable coupling
model as long as the initial entanglement is maximal.  For other initial states
the disagreement is progressive as the dimension of the environment increases.
We conjecture that in the limit of large $m$ there will be a dynamical quantum
phase transition between good agreement and maximum disagreement, for
arbitrarily small deviations from a Bell state.  \par 
\section{Other functions: von Neumann entropy} \label{sec:entropy} 
\begin{figure} 
\begin{center}
\includegraphics[scale=1.4]{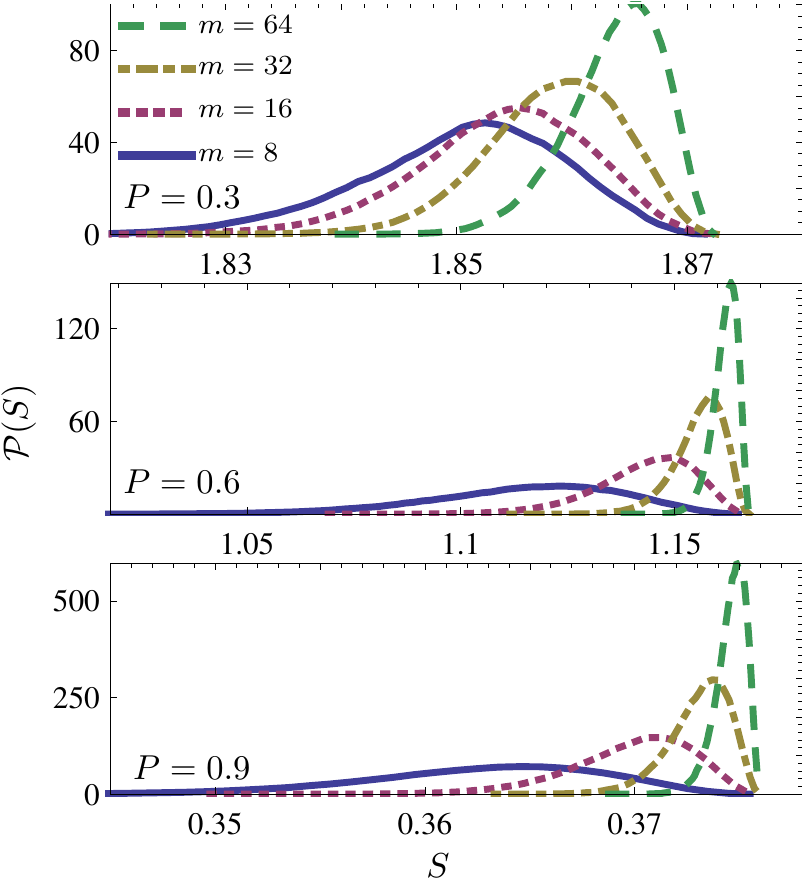}
\end{center}
\caption{\label{fig:entropy:for:purity} (Color online)  Probability
distribution of the entropy $S$ for several values of purity. Each plot shows a
particular purity and several dimensions of the environment.  Notice the scale
in the horizontal axis to observe that the distributions are already quite
peaked for the dimensions shown, the effect being larger for increasing dimension. }
\end{figure} 
We have also studied the von Neumann entropy (henceforth called entropy) of the reduced density 
matrix
\begin{equation}
S(\rho) = -\tr \left( \rho \ln \rho \right). 
\label{eq:vonneumann}
\end{equation}
The algebraic structure is considerably more complicated than for purity, and
thus such a detailed program as was presented in the previous sections is in general only
possible in terms of solutions of transcendental equations. However much can be
said using the fact that purity and entropy are closely related in typical
cases. 

For the ensembles studied in \sref{sec:static}, consider the entropy. 
Its distribution is plotted in \fref{fig:entropy:for:purity}, for several fixed
purities, and several dimensions of the environment. 
One can see that even for small dimensions of the environment ($m=8$), the
width of the entropy distribution is already of order $10^{-2}$, and its value
decreases with increasing $m$.  Thus, even though the ensembles for fixed
purity and fixed entropy are different (not even the support is the same), they
are similar. 

As an example, the equivalent to figures \ref{fig:example:smalldims}
and \ref{fig:manchatwoD}, but for fixed entropy,
is shown in \fref{fig:example:entropy}.
We obtained numerically the corresponding limiting curves and nodes, and the
probability distribution corresponding to the static ensemble. The resemblance
to the case of purity is striking, and one may conclude that many of the
results obtained for a fixed value of purity must hold qualitatively  for a
fixed value of the entropy.

Notice that we are considering just two functions of infinitely many that define
a partial order over the eigenvalue vector, which mathematically corresponds to
a probability distribution. This indicates that the entire theory developed by
Hardy, Littlewood, and
Polya~\cite{hardy1952inequalities} applies, as noted in
reference~\cite{ru1,ru2, Uhlmann}. In particular it indicates that the ordering  becomes
unique both near the pure states and near the maximally mixed state, thus 
formalizing the analogy between purity and, say, entropy when one approaches 
any of these limits. 


\begin{figure} 
\begin{center}
\includegraphics[width=\columnwidth]{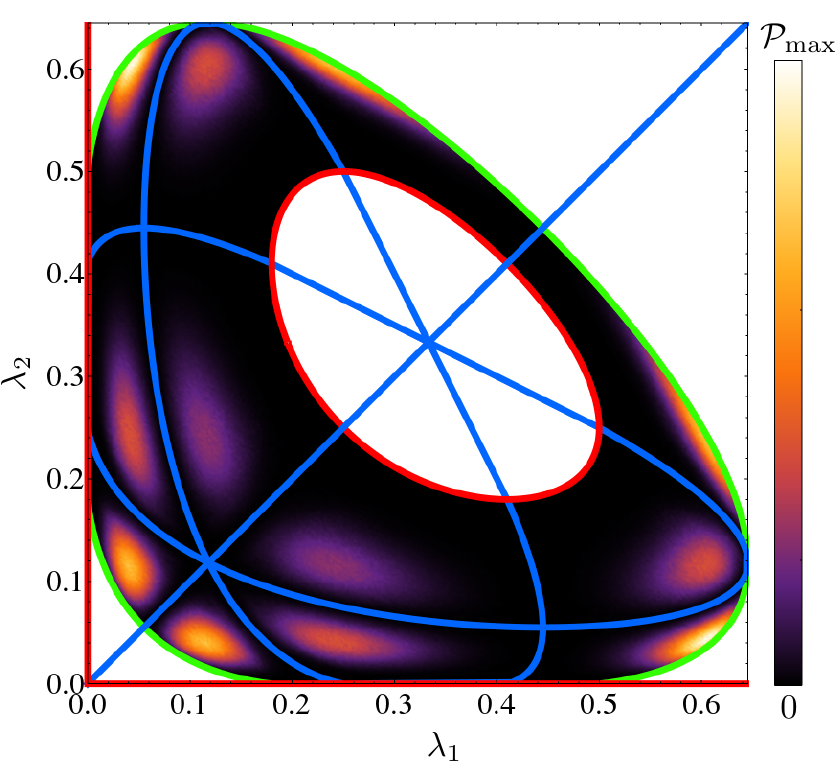}
\end{center}
\caption{\label{fig:example:entropy} (Color online)  Probability distribution
of 2 eigenvalues at fixed entropy $S=1.5$ for $m=8$, in arbitrary units. The
white region leads to nonphysical eigenvalues. Nodes and limiting curves are
also present here. The color coding is exactly the same as in figures \ref{fig:example:smalldims} and 
\ref{fig:manchatwoD}. }
\end{figure} 
\section{Conclusions} 

In this paper we set out to give a random matrix model for density matrices of
open quantum system by imposing further restrictions on an ensemble
originally obtained by simply restricting an ensemble of random covariance
matrices to matrices with unit trace.  In section two we give some general
remarks about the alternate possibilities of restricting random matrix
ensembles in analogy to techniques used in statistical mechanics, as well as on
the alternative of a proposing a random matrix description directly for the
object of interest, as compared to proposing a random set of Hamiltonians to
describe the dynamics and from that calculate the ensemble of that same object.
In the latter case the condition is imposed by following evolution until it is
fulfilled exactly (i.e. up to numerical exactitude).
 
The conditions can either be imposed in a ``strict way'' for each member of the
ensemble resulting in microcanonical ensembles or on average in the spirit of
Jayne's principle  leading to canonical ensembles. In view of the original
construction \cite{Nadal2011} we opt for the microcanonical approach.  

After the construction of the ensemble we concentrate on the distribution of
eigenvalues of the density matrices using fixed purity as the additional
restriction. The choice of purity allows us to obtain analytical results but we
also consider alternate convex functions and explicitly the von Neumann entropy
towards the end of the paper. Qualitatively very similar rests are obtained  

To visualize our result we restrict our considerations to a four dimensional
central system. The two restrictions imposed allow to give density maps over
the two remaining dimensions and one clearly sees the signature of the Jacobi
determinants on an otherwise smooth distribution limited by the boundaries
imposed by the restrictions on trace and purity. Again, this picture remains
qualitatively unchanged if we replace purity by von Neumann entropy.

For this four dimensional central system we present the comparison with
the results obtained if we run the time evolution of an initially pure state of
the central system under the random matrix model for decoherence given in terms
of several families of random Hamiltonians for the full system (central system
+ environment). We find fairly good agreement if the fouling between
environment and the central
system is not too weak, i.e. if we are not in the ``perturbative regime''.
Surprisingly this holds true even if the four dimensional central system
consists of two quits but one of the two does not interact with either the
other or the environment, under the condition that the two qubits are maximally
entangled.


We may speculate as to why the static and dynamic models agree so well, except
for the case of partially entangled central systems with a structured coupling
to the environment.  Consider POSETS (partially ordered sets or lattices) of
density matrices introduced by Uhlmann~\cite{Uhlmann} as well as by Ruch and Mead
\cite{ru1, ru2} independently.
This work is based on a partial order introduced by Muirhead~\cite{muirhead} and connected to
classical bi-stochastic evolution by Hardy, Littlewood and
Polya~\cite{hardy1952inequalities} (see also
\cite{ru1}). The classical evolution can be generalized  to POSETS of pairs of
distribution~\cite{HLP1,HLP2} and such concepts seem to have a growing impact on recent
work using majorization in the context of open quantum systems (e.g 
\cite{wilming2014weak, faist2015gibbs, lostaglio2015quantum, gour2015resource}).

In this context it is well known, that near distributions with a dominant
component (i.e. for almost pure states) all convex functions will yield
essentially the same order, or, with other words, near its extreme points
POSETS are almost completely ordered sets.  Therefore a unitary dynamic acting
on a initially separable density matrix will have no choice but to follow
closely the time-independent partial order, which governs the static ensemble.
Indeed if we look at \fref{fig:global:hamiltonian} we see immediately that the
agreement is best near maximal purity.

A number of steps along these lines are possible for the future. One may
investigate if a 
partial orders for pairs, as suggested above, can be extended to pairs of density matrices.
is feasible and weather it it brings any
advantages. 
A note of caution must be added here, as the partial order for pairs of
distributions in \cite{HLP1, HLP2} has not been successfully generalized to density
matrices because the two members of the pair typically cannot be diagonalized
with the same transformation.
If and when this mathematical problem is overcome we may well find this to be
the key to a deeper understanding of our results.
Another more obvious next step is to calculate other quantities for
fixed decoherence as measured by purity or some other convex function. For the
two qubit case concurrence would be the obvious candidate, as it would allow to
calculate $CP$ diagrams (concurrence purity) without referring to time
evolution. 

\ack 
We thank Satya Majumdar for very stimulating discussions at the onset of this project, 
and Francois Leyvraz for informative discussions near its culmination. 
We thank Isaac Perez for help with the Monte Carlo simulations.
Support by the  projects CONACyT 153190, 79613, 219993, 
UNAM-PAPIIT IA101713, IN111015 
and IG101113 are acknowledged.
\section*{References} 
\bibliographystyle{unsrt}
\bibliography{paperdef,mibibliografia,library,local}

\begin{thebibliography}{10}

\bibitem{vN55a}
J.~von Neumann.
\newblock {\em Mathematical Foundations of Quantum Mechanics}.
\newblock Princeton University Press, 1955.

\bibitem{breuer2007theory}
H.P. Breuer and F.~Petruccione.
\newblock {\em The Theory of Open Quantum Systems}.
\newblock OUP Oxford, 2007.

\bibitem{Lindblad1976}
G.~Lindblad.
\newblock {On the generators of quantum dynamical semigroups}.
\newblock {\em Communications in Mathematical Physics}, 48(2):119--130, June
  1976.

\bibitem{Gorini1976}
Vittorio Gorini, Andrzej Kossakowski, and E~C~G Sudarshan.
\newblock {Completely positive dynamical semigroups of N-level systems}.
\newblock {\em Journal of Mathematical Physics}, 17(5):821, May 1976.

\bibitem{Gorin2003}
T.~Gorin and T.~H. Seligman.
\newblock Decoherence in chaotic and integrable systems: a random matrix
  approach.
\newblock {\em Phys. Lett. A}, 309:61--67, Mar 2003.

\bibitem{2008NJPh...10k5016G}
T.~{Gorin}, C.~{Pineda}, H.~{Kohler}, and T.~H. {Seligman}.
\newblock {A random matrix theory of decoherence}.
\newblock {\em New J. Phys.}, 10(11):115016--+, November 2008.

\bibitem{PhysRevLett.104.110501}
C.~Nadal, S.~N. Majumdar, and M.~Vergassola.
\newblock Phase transitions in the distribution of bipartite entanglement of a
  random pure state.
\newblock {\em Phys. Rev. Lett.}, 104(11):110501, Mar 2010.

\bibitem{Nadal2011}
Celine Nadal, Satya~N. Majumdar, and Massimo Vergassola.
\newblock {Statistical Distribution of Quantum Entanglement for a Random
  Bipartite State}.
\newblock {\em Journal of Statistical Physics}, 142(2):403--438, January 2011.

\bibitem{Mello1985254}
P.~A. Mello, P.~Pereyra, and T.~H. Seligman.
\newblock Information theory and statistical nuclear reactions. i. general
  theory and applications to few-channel problems.
\newblock {\em Annals of Physics}, 161(2):254 -- 275, 1985.

\bibitem{haakebook}
F.~Haake.
\newblock {\em Quantum Signatures of Chaos, II ed.}
\newblock Springer, Berlin, 2001.

\bibitem{pinedalong}
C.~Pineda, T.~Gorin, and T.~H. Seligman.
\newblock Decoherence of two-qubit systems: a random matrix description.
\newblock {\em New J. Phys.}, 9(4):106, April 2007.

\bibitem{PhysRevLett.107.080404}
M.~\ifmmode \check{Z}\else \v{Z}\fi{}nidari\ifmmode~\check{c}\else \v{c}\fi{},
  C.~Pineda, and I.~Garc\'\i{}a-Mata.
\newblock Non-markovian behavior of small and large complex quantum systems.
\newblock {\em Phys. Rev. Lett.}, 107(8):080404, Aug 2011.

\bibitem{pinedaRMTshort}
C.~Pineda and T.~H. Seligman.
\newblock Bell pair in a generic random matrix environment.
\newblock {\em Phys. Rev. A}, 75(1):012106, 2007.

\bibitem{Agassi1975145}
D~Agassi, H.A Weidenmüller, and G~Mantzouranis.
\newblock The statistical theory of nuclear reactions for strongly overlapping
  resonances as a theory of transport phenomena.
\newblock {\em Physics Reports}, 22(3):145 -- 179, 1975.

\bibitem{VWZ}
J.~J.~M. Verbaarschot, H.~A Weidenm{\"u}ller, and M.~R. Zirnbauer.
\newblock Grassmann integration in stochastic quantum physics: the case of
  compound-nucleus scattering.
\newblock {\em Phys. Rev.}, 129(6):367--438, 1985.

\bibitem{SCali}
T.~H. Seligman.
\newblock Analitic and constructive use of information theory in physics.
\newblock In S.~Moore and B.G. Moreno, editors, {\em Stochastic Processes
  Applied to Physics and Other Related Fields: Cali, Colombia, 21 June - 9 July
  1982}, Singapore, 1983. World Scientific.

\bibitem{cartanRMT}
{\'E}.~Cartan.
\newblock Quasi composition algebras.
\newblock {\em Abh. Math. Sem. Hamburg}, 11:116, 1935.

\bibitem{mehta}
M.~L. Mehta.
\newblock {\em Random Matrices}.
\newblock Academic Press, San Diego, California, second edition, 1991.

\bibitem{bohr1939mechanism}
N.~Bohr and J.~A. Wheeler.
\newblock The mechanism of nuclear fission.
\newblock {\em Phys. Rev.}, 56(5):426, 1939.

\bibitem{pineda:012305}
C.~Pineda and T.~H. Seligman.
\newblock Evolution of pairwise entanglement in a coupled $n$-body system.
\newblock {\em Phys. Rev. A}, 73(1):012305, 2006.

\bibitem{gorin2008random}
T.~Gorin, C.~Pineda, H.~Kohler, and T.H. Seligman.
\newblock A random matrix theory of decoherence.
\newblock {\em New J. Phys.}, 10(11):115016, 2008.

\bibitem{MGS2015}
H.~J. Moreno, T.~Gorin, and T.~H. Seligman.
\newblock Improving coherence with nested environments.
\newblock {\em Phys. Rev. A}, 92:030104, Sep 2015.

\bibitem{jaynes1}
E.~T. Jaynes.
\newblock Information theory and statistical mechanics. ii.
\newblock {\em Phys. Rev.}, 108:171--190, Oct 1957.

\bibitem{jaynes2}
E.~T. Jaynes.
\newblock Information theory and statistical mechanics.
\newblock {\em Phys. Rev.}, 106:620--630, May 1957.

\bibitem{rlevine87:dynamics}
Raphael~D. Levine and Richard~B. Bernstein.
\newblock {\em Molecular Reaction Dynamics and Chemical Reactivity}.
\newblock Oxford University Press, Inc., New York, 1987.

\bibitem{SM2Nuclphys}
P.A. Mello and T.H. Seligman.
\newblock On the entropy approach to statistical nuclear reactions.
\newblock {\em Nucl. Phys. A}, 344(3):489--508, 1980.

\bibitem{SM31chann}
P.A. Mello.
\newblock A statistical theory of nuclear reactions based on a variational
  principle.
\newblock {\em Phys. Lett. B}, 81(2):103 -- 106, 1979.

\bibitem{baranger1994mesoscopic}
H.~U. Baranger and P.~A. Mello.
\newblock Mesoscopic transport through chaotic cavities: A random s-matrix
  theory approach.
\newblock {\em Phys. Rev. Lett.}, 73(1):142, 1994.

\bibitem{brouwer1995generalized}
P.~W. Brouwer.
\newblock Generalized circular ensemble of scattering matrices for a chaotic
  cavity with nonideal leads.
\newblock {\em Phys. Rev. B}, 51(23):16878, 1995.

\bibitem{kuhl2005direct}
U.~Kuhl, M.~Mart{\'\i}nez-Mares, R.A. M{\'e}ndez-S{\'a}nchez, and H.-J.
  St{\"o}ckmann.
\newblock Direct processes in chaotic microwave cavities in the presence of
  absorption.
\newblock {\em Phys. Rev. Lett.}, 94(14):144101, 2005.

\bibitem{wishartoriginal}
Seth Lloyd and Heinz Pagels.
\newblock Complexity as thermodynamic depth.
\newblock {\em Ann. Phys.}, 188(1):186 -- 213, 1988.

\bibitem{1464-4266-4-4-325}
T.~Gorin and T.~H. Seligman.
\newblock A random matrix approach to decoherence.
\newblock {\em J. Opt. B}, 4(4):S386, 2002.

\bibitem{PhysRevA.90.022107}
M.~Carrera, T.~Gorin, and T.~H. Seligman.
\newblock Single-qubit decoherence under a separable coupling to a random
  matrix environment.
\newblock {\em Phys. Rev. A}, 90:022107, Aug 2014.

\bibitem{hardy1952inequalities}
G.H. Hardy, J.E. Littlewood, and G.~P{\'o}lya.
\newblock {\em Inequalities}.
\newblock Cambridge Mathematical Library. Cambridge University Press, 1952.

\bibitem{ru1}
E.~Ruch.
\newblock The diagram lattice as structural principle a. new aspects for
  representations and group algebra of the symmetric group b. definition of
  classification character, mixing character, statistical order, statistical
  disorder; a general principle for the time evolution of irreversible
  processes.
\newblock {\em Theor. Chim. Acta}, 38(3):167--183, 1975.

\bibitem{ru2}
Ernst Ruch and Alden Mead.
\newblock The principle of increasing mixing character and some of its
  consequences.
\newblock {\em Theor. Chim. Acta}, 41(2):95--117, 1976.

\bibitem{Uhlmann}
A.~Uhlmann.
\newblock Endlich-dimensionale dichtematrizen. i.
\newblock {\em Wissenschaftliche Zeitschrift der Karl-Marx-Universit{\"a}t
  Leipzig. Mathematisch-Naturwissenschaftliche Reihe}, 21:421--452, 1972.

\bibitem{muirhead}
R.~F. Muirhead.
\newblock Some methods applicable to identities and inequalities of symmetric
  algebraic functions of n letters.
\newblock {\em Proceedings of the Edinburgh Mathematical Society}, 21:144--162,
  2 1902.

\bibitem{HLP1}
E.~Ruch, R.~Schranner, and T.~H. Seligman.
\newblock The mixing distance.
\newblock {\em J. Chem. Phys.}, 69(1):386--392, 1978.

\bibitem{HLP2}
E.~Ruch, R.~Schranner, and T.~H. Seligman.
\newblock {Generalization of a theorem by Hardy, Littlewood, and Pólya}.
\newblock {\em J. Math. Anal. Appl.}, 76(1):222 -- 229, 1980.

\bibitem{wilming2014weak}
H.~Wilming, R.~Gallego, and J.~Eisert.
\newblock Weak thermal contact is not universal for work extraction.
\newblock {\em arXiv preprint arXiv:1411.3754}, 2014.

\bibitem{faist2015gibbs}
P.~Faist, J.~Oppenheim, and R.~Renner.
\newblock Gibbs-preserving maps outperform thermal operations in the quantum
  regime.
\newblock {\em New J. Phys.}, 17(4):043003, 2015.

\bibitem{lostaglio2015quantum}
M.~Lostaglio, K.~Korzekwa, D.~Jennings, and T.~Rudolph.
\newblock Quantum coherence, time-translation symmetry, and thermodynamics.
\newblock {\em Phys. Rev. X}, 5(2):021001, 2015.

\bibitem{gour2015resource}
G.~Gour, M.~P M{\"u}ller, V.~Narasimhachar, R.~W Spekkens, and N.~Y. Halpern.
\newblock The resource theory of informational nonequilibrium in
  thermodynamics.
\newblock {\em Phys. Rep.}, 583:1 -- 58, 2015.

\end{thebibliography}
\appendix
\section{A isometrical description for the 2 qubit case} 
\begin{figure} 
\begin{center}
\includegraphics{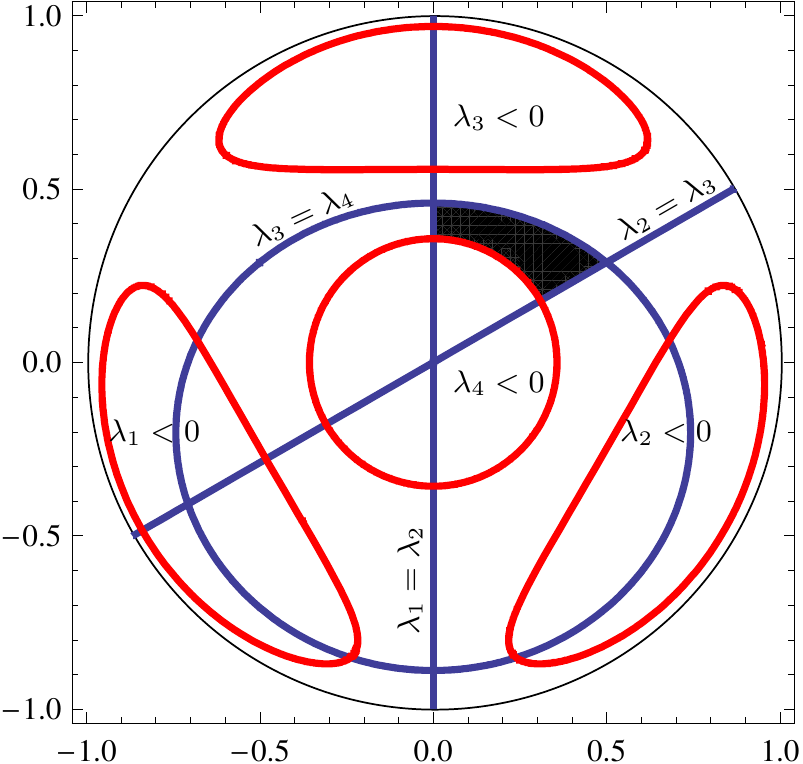}
\end{center}
\caption{\label{fig:twoD} (Color online)  Region of interest. The 
boundary between positive (physical) eigenvalues and negative is showed in 
red, for each of the eigenvalues. The border between some degenerate 
eigenvalues is plotted in blue. Thus, the physical region with 
sorted eigenvalues restricts only to the black area. }
\end{figure} 
For the 2 qubit case, one case visualize the distribution of eigenvalues in the 
$(\lambda_1,\lambda_2)$ plane. However this is not the only way of visualizing the 
results. One would like a way that preserves the natural symmetry of the problem, 
and that conserves a notion of ``volume'' in this space. 
Taking the natural volume in $\Re^n$, one can come up with an isometry, after
rotating, doing an isometrical stereographical projection, a simple plane.
There, one can calculate the position of triple degenerate
states, and the physical regions ($\lambda_i>0$). 
\par
One reason to perform the analysis done with the other visualization, is that 
reading the eigenvalues directly from a point in the plane is straightforward. With 
the representation presented in this appendix, it is not the case. One thus sacrifices 
symmetry and isometry for clarity. 
\par
We shall initially restrict to the 3 dimensional space defined by the
normalization condition. We found convenient to transform condition
\ref{eq:normalization} to one that involves a single variable. A rotation seems
the most appropriate, and since the distance to the origin must remain
invariant, one has to search for a rotation that transforms \ref{eq:normalization} to 
\begin{equation}
\lambda_1 = 1/2.
\label{eq:simple_hyperplane}
\end{equation}
This is the same rotation that transforms the vector $(1/2, 1/2, 1/2, 1/2)$ (normal to the 
hyperplane   \ref{eq:normalization}) to the vector $(1,0,0,0)$ (normal to the 
hyperplane \ref{eq:simple_hyperplane}). One can build such a rotation with a sequence
of 2-dimensional rotations, the first acting on the last two coordinates and transforming
$(1/2,1/2)$ to $(\sqrt 2/2,0)$, the second one acting on the second and third coordinate 
and transforming thus $(1/2, \sqrt 2/2)$ to $(\sqrt 3 /2,0 )$ and the last one 
acting on the first two coordinates and taking $(1/2, \sqrt 3 /2 )$ to 
$(1,0)$. The overall sequence of rotations lead to the matrix 
\begin{equation}
R = \begin{pmatrix}
\frac{1}{\sqrt 2} & -\frac{1}{\sqrt 2} & 0&0 \\
\frac{1}{\sqrt 6} & \frac{1}{\sqrt 6} & -\frac{\sqrt 6}{3} & 0 \\
\frac{1}{2 \sqrt 3} & 
\frac{1}{2 \sqrt 3} & 
\frac{1}{2 \sqrt 3} & 
 -\frac{\sqrt 3}{2}  \\
 1/2 & 1/2 & 1/2 & 1/2 
\end{pmatrix}.
\label{eq:rotationmatrix}
\end{equation}
The condition regarding purity transforms trivially: a rotated hyper-sphere is
a hyper-sphere.  The intersection of the ``rotated'' sphere with the rotated
hyperplane \eref{eq:simple_hyperplane} can be readily evaluated, and yields
\begin{equation}
\sum_{i=2}^4\lambda_i^2 = P -\frac14
\label{eq:purity_restricted}
\end{equation}
which is the familiar 3-sphere with a radius ranging from 0 to $3/4$.  

The planes which describe the semi positivity condition of the eigenvalues are transformed
into planes in $\mathbb R^3$ with equations 
\begin{subequations}
\begin{align}
2\sqrt{3}\lambda_4 &\le 1 \\
2\sqrt{2}\lambda_3-\lambda_4 &\le \frac{\sqrt{3}}{2} \\
\sqrt{6}\lambda_2 - \sqrt{2}\lambda_3-\lambda_4 &\le \frac{\sqrt{3}}{2} \\
-\sqrt{6}\lambda_2 - \sqrt{2}\lambda_3-\lambda_4 &\le \frac{\sqrt{3}}{2} 
\label{eq:rotated_semipositive}
\end{align}
\end{subequations}
These planes define a regular tetrahedron, inside which must lay the
(rotated) eigenvalues of any physical density matrix. 
For $P<1/3$ the sphere is completely inside the tetrahedron, and at
$P=1/3$ it is tangent to the faces. At $P=1/2$ the sphere touches the 
vertexes. Finally, at $P=1$ it touches the tetrahedron only in the corners, allowing
only 4 situations, namely when a single eigenvalue is 1 and the others are
0.  We invite the reader to refer to Figs.~\ref{fig:threeandtwoD}, (a) and (c).

The next step is  to project the sphere obtained to a plane, but keeping the
natural measure in $\mathbb R^n$. Thus the usual stereographic projection is not an
option.  We write the point in the three dimensional space, previously found, 
in spherical coordinates $(r,\theta,\phi)$, and map
them to polar coordinates in the plane $(R,\vartheta)$. As for a fixed purity
one has constant $r$, this coordinate can be ignored. Mapping $\phi \to
\vartheta$ and choosing $R$ as a function (to be determined) of $\theta$ so as to make an ``stretched''
stereographic projection, will do the required job. With that freedom one can make
the transformation an isometry. The isometry requirement can 
be written formally as 
\begin{equation} 
\sin \theta \rmd \theta \propto R \rmd R 
\label{eq:condition_isometry} 
\end{equation} 
ignoring a normalization constant. This can be fulfilled using $R = \sin (\theta/2)$. 

In \fref{fig:twoD}, we present the limiting regions studied before. That is, the regions
which delimit the semi-positive definitiveness of the density matrix. That is, where 
the eigenvalues are zero. This results in four red curves, with a threefold symmetry. 
Moreover, the curves that represent the degenerate points ($\lambda_i = \lambda_j$) necessary 
to delimit the region that leads to ordered eigenvalues are presented in blue. That region
is colored in black. 

We are now interested in the allowed areas [black regions in 
\fref{fig:threeandtwoD} (b) and (d)]. However an analytic form of the
curve seems to be too complicated to be obtained. One can still 
appreciate a high degree of symmetry in the plot. This is due to the
possibility of exchanging different eigenvalues. This brings up 24 equivalent
zones, since there are $4! = 24$ ways of ordering 4 different objects.  Since
the ordering will not bring any difference in the distributions, one can restrict
to the area in which
\begin{equation}
\lambda_i \ge \lambda_{i+1}.
\label{eq:order_eigenvalues}
\end{equation}

\begin{figure} 
\begin{center}
\includegraphics[viewport=58 460 300 750,clip, scale=1.4]{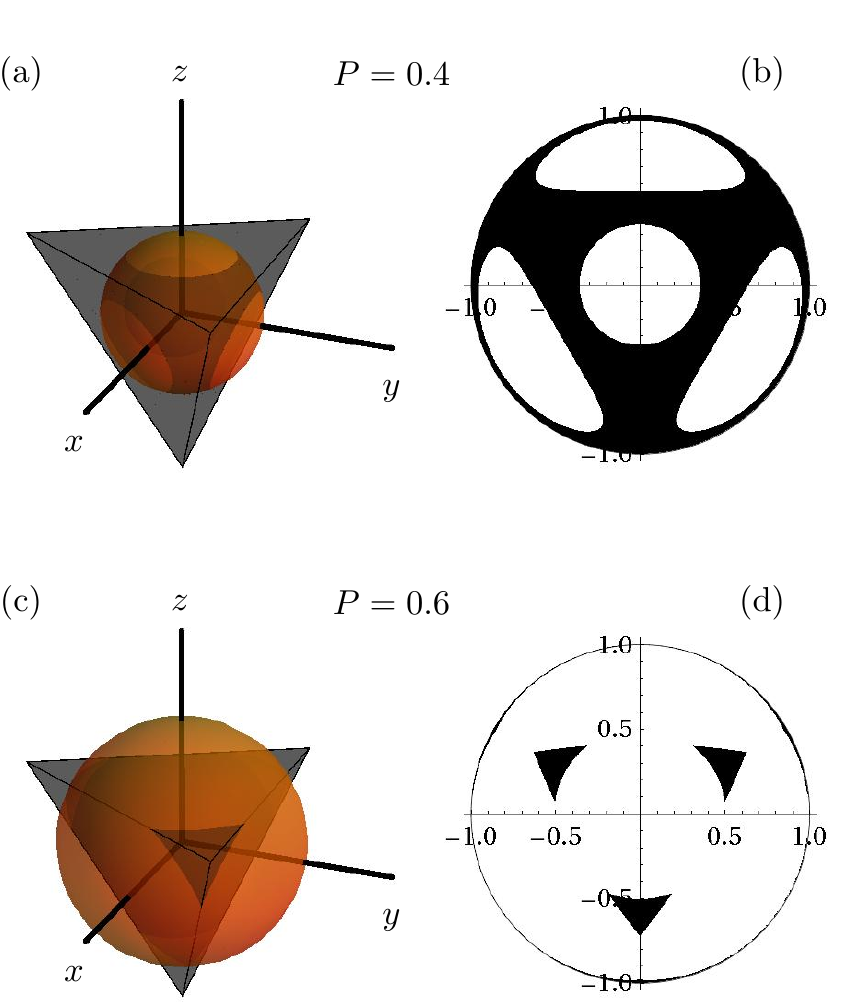}
\end{center}
\caption{\label{fig:threeandtwoD} (Color online)  Representation of the 
eigenvalues in the hyperplane where the normalization condition is 
met (left) and its projection to two dimensions using a measure preserving map. 
We show two values of purity ($P=0.4,0.6$ for the top/bottom figure) where
the region is connected and disconnected.  }
\end{figure} 
\section{Distribution of individual eigenvalues}  
\label{sec:distribution:smallest:eigenvalues}

One can consider the individual distributions of the ordered eigenvalues. We
can study such distribution in the $(\lambda_1, \lambda_2)$ plane. We shall
then restrict to the area that contains $\lambda_i\ge\lambda_{i+1}$ indicated
by a dark curve enclosing the area of interest in \fref{fig:symmetries}. 
To obtain the distribution of the smallest eigenvalue, 
the following integral must be performed:
\begin{multline}
\mcP_{\lambda_1}(\lambda_{1, \min}\le\lambda_1\le \lambda_{1, \max}) \\ \propto 
\int_{\lambda_{2, \min}(\lambda_1)}^{\lambda_{2, \max}(\lambda_1)}
\rmd \lambda_2
\prod_i \lambda_i^{m-4} \prod_{i<j} (\lambda_i-\lambda_j)^2
\label{eq:lambdaone}
\end{multline}
with $\lambda_{3,4}=\lambda_{3,4}(\lambda_1, \lambda_2)$ given by 
\eref{eq:lambdapm} and the limits being given by the relations
\begin{align}
\lambda_{1, \min} &= \max \left\{0, \frac14 \left( 1-\sqrt{12P -3} \right) \right\} \\
\lambda_{1, \max} &=\frac{3-\sqrt{12P -3}}{12} \\
\lambda_{2, \min}(\lambda_1) &= \max \left\{\lambda_1, \frac{1-\lambda_1-\sqrt{6P-2+4\lambda_1-8\lambda_1^2} }{3} \right\}\\
\lambda_{2, \max}(\lambda_1) &= \frac{2-2\lambda_1-\sqrt{6P-2+4\lambda_1-8\lambda_1^2} }{6}.
\label{eq:limitsone}
\end{align}
Similarly one can obtain the distribution for the second smallest eigenvalue:
\begin{multline}
\mcP_{\lambda_2}(\lambda_{2, \min}\le\lambda_2\le \lambda_{2, \max}) \\ 
\propto \int_{\lambda_{1, \min}(\lambda_2)}^{\lambda_{1, \max}(\lambda_2)}
\rmd \lambda_1
\prod_i \lambda_i^{m-4} \prod_{i<j} (\lambda_i-\lambda_j)^2
\label{eq:lambdatwo}
\end{multline}
with 
\begin{align}
\lambda_{2, \min} &= \max \left\{0, \frac14 \left( 1-\sqrt{4P -1} \right) \right\} \\
\lambda_{2, \max} &=
\begin{cases}
\frac{2-\sqrt{6P -2}}{6} &\text{$P>1/3$}, \\
\frac{3-\sqrt{12P -3}}{12} &\text{$P\le1/3$}
\end{cases} \\
\lambda_{1, \min}(\lambda_2) &= \max \left\{0, \frac{1-\lambda_2-\sqrt{6P-2+4\lambda_2-8\lambda_2^2} }{3} \right\}\\
\lambda_{1, \max}(\lambda_2) &= \min \left\{\lambda_2, \frac{1-2\lambda_2-\sqrt{2P-1+4\lambda_2-8\lambda_2^2} }{2} \right\}.
\label{eq:limitstwo}
\end{align}
The expressions for the projected probabilities are straightforward to obtain 
for fixed $m$,
but in general result in lengthy expressions, that can be obtained from a computer
program. 	

\section{Monte Carlo simulations} 
\label{sec:montecarlo}
We performed Monte Carlo simulations of the eigenvalues at fixed
purity and entropy. This was done for the following three reasons. (i)
To check the formulae obtained; (ii) to estimate the statistical error 
in calculating the Kolmogorov distance, due both to finite sampling 
and finite binning; and (iii) to calculate the distribution of the eigenvalues 
at fixed entropy, as it is easier than solving the resulting transcendental
equation. Inspired in 
\cite{Nadal2011}, we consider a potential energy for the eigenvalues of 
\begin{equation}
E(\vec \lambda) = 
-(m-4) \sum_i \log \lambda_i - 2 \sum_{i < j} \log |\lambda_i - \lambda_j| 
\label{eq:potential:monte:carlo}
\end{equation}
for $ \lambda_i \in \mathbb{R}^+$.
A step in which the first two eigenvalues are displaced in a random
angle a distance $\epsilon$ is used, and the other two are obtained by requiring
normalization and the desired value of purity. That way, fixed purity and normalization 
condition are ensured. 
Moreover, if this results in a negative
eigenvalue, its absolute value is taken. With these conditions, 
$\epsilon$ is set such that the acceptance rate is between $0.4$ and $0.6$. The
particular value depends strongly on both $m$ and $P$. To perform the simulations
for fixed entropy, one can simply solve for the other two eigenvalues, but, of course,
at fixed entropy. This is however considerably more demanding.
This method can be generalized for arbitrary dimensions, one should just perform a random walk
in $n-2$ dimensions, with potential energy given by eq.~\eref{eq:potential:monte:carlo}, and 
calculate the other two eigenvalues to fulfill the conditions of normalization and fixed, say,
purity. 
\end{document}
"VWZ"
"SM2Nuclphys"
"SM31chann"
"TBP"
"Uhlmann"